\documentclass[%
 reprint,
superscriptaddress,
 amsmath,amssymb,
 aps,
prb,
floatfix,
]{revtex4-1}

\usepackage{graphicx}% Include figure files
\usepackage{dcolumn}% Align table columns on decimal point
\usepackage{bm}% bold math
\usepackage[colorlinks=true,allcolors=blue]{hyperref}% add hypertext capabilities
\usepackage{array,booktabs,tabularx}
\usepackage[caption=false]{subfig}
\usepackage[USenglish]{babel}
\usepackage{braket}

\begin{document}
\selectlanguage{USenglish}
\preprint{APS/123-QED}

\title{Strong anharmonicity and high thermoelectric efficiency in high temperature SnS from first-principles}

\author{Unai Aseginolaza}
%\email{uaseguinolaz001@ehu.eus}
\affiliation{Centro de F\'isica de Materiales CFM, CSIC-UPV/EHU, Paseo Manuel de
             Lardizabal 5, 20018 Donostia, Basque Country, Spain}
\affiliation{Donostia International Physics Center
             (DIPC), Manuel Lardizabal pasealekua 4, 20018 Donostia, Basque Country, Spain}
\affiliation{Fisika Aplikatua 1 Saila, Gipuzkoako Ingenieritza Eskola,
             University of the Basque Country (UPV/EHU), Europa Plaza 1, 20018 Donostia,
             Basque Country, Spain}
\author{Raffaello Bianco}
\affiliation{Centro de F\'isica de Materiales CFM, CSIC-UPV/EHU, Paseo Manuel de
             Lardizabal 5, 20018 Donostia, Basque Country, Spain}
\affiliation{Department of Applied Physics and Material Science, Steele Laboratory, California Institute of Technology, Pasadena,
California 91125, United States}
\author{Lorenzo Monacelli}
\affiliation{Dipartimento di Fisica, Universit\`a di Roma La Sapienza, Piazzale Aldo Moro 5, I-00185 Roma, Italy}
\author{Lorenzo Paulatto}
\affiliation{IMPMC, UMR CNRS 7590, Sorbonne
Universit\'es - UPMC Univ. Paris 06, MNHN, IRD, 4 Place Jussieu,
F-75005 Paris, France}
\author{Matteo Calandra}
\affiliation{Sorbonne Universit\'es, CNRS, Institut des Nanosciences de Paris, UMR7588, F-75252, Paris, France}
\author{Francesco Mauri}
\affiliation{Dipartimento di Fisica, Universit\`a di Roma La Sapienza, Piazzale Aldo Moro 5, I-00185 Roma, Italy} 
\affiliation{Graphene Labs, Fondazione Instituto Italiano di Tecnologia, Italy}
\author{Aitor Bergara}
%\email{a.bergara@ehu.eus}
\affiliation{Centro de F\'isica de Materiales CFM, CSIC-UPV/EHU, Paseo Manuel de
             Lardizabal 5, 20018 Donostia, Basque Country, Spain}
\affiliation{Donostia International Physics Center
             (DIPC), Manuel Lardizabal pasealekua 4, 20018 Donostia, Basque Country, Spain}
\affiliation{Departamento de F\'isica de la Materia Condensada,  University of the Basque Country (UPV/EHU), 48080 Bilbao, 
             Basque Country, Spain}
\author{Ion Errea}
%\email{ion.errea@ehu.eus}
\affiliation{Centro de F\'isica de Materiales CFM, CSIC-UPV/EHU, Paseo Manuel de
             Lardizabal 5, 20018 Donostia, Basque Country, Spain}
\affiliation{Donostia International Physics Center
             (DIPC), Manuel Lardizabal pasealekua 4, 20018 Donostia, Basque Country, Spain}
\affiliation{Fisika Aplikatua 1 Saila, Gipuzkoako Ingenieritza Eskola,
             University of the Basque Country (UPV/EHU), Europa Plaza 1, 20018 Donostia,
             Basque Country, Spain}

\date{\today}% It is always \today, today,

\begin{abstract}
SnS and SnSe are isoelectronic materials with a common phase diagram. Recently, SnSe was found to be the most efficient intrinsic thermoelectric material in its high-temperature $Cmcm$ phase above $800$ K. Making use 
of first-principles calculations, here we show that the electronic and vibrational properties of both materials are very similar in this phase and, consequently, SnS is also expected to have a high thermoelectric figure 
of merit at high temperature in its $Cmcm$ phase. In fact, the electronic power factor and lattice thermal conductivity are comparable for both materials, which ensures a similar figure of merit. As in the case of SnSe, the 
vibrational properties of SnS in the $Cmcm$ phase are far from trivial and are dominated by huge anharmonic effects. Its phonon spectra are strongly renormalized by anharmonicity and the spectral functions of some particular 
in-plane modes depict anomalous non-lorentzian profiles. Finally, we show that non-perturbative anharmonic effects in the third-order force-constants are crucial in the calculation of the lattice thermal conductivity. Our results 
motivate new experiments in the high temperature regime to measure the figure of merit of SnS.
\end{abstract}

                             % Classification Scheme./
\maketitle

\section{Introduction}

Thermoelectricity is a technologically  interesting material property that allows to transform residual heat into useful electricity\cite{goldsmid2010introduction,behnia2015fundamentals}. The efficiency of this energy transformation 
is controlled by the dimensionless figure of merit 
\begin{equation}
ZT=S^{2}\sigma T/\kappa, 
\end{equation}
where $S$ is the Seebeck coefficient, $\sigma$ the electrical conductivity, $T$ the temperature, and $\kappa=\kappa_{e}+\kappa_{l}$ the sum of electronic $\kappa_{e}$ and lattice $\kappa_{l}$ thermal conductivities. Therefore, a 
good thermoelectric performance requires a high power factor $P_{F}=S^{2}\sigma$ together with a low thermal conductivity. 

Monochalcogenides have proven to be efficient thermoelectric materials\cite{heremans2008enhancement,zhang2013high,yang2008nanostructures,cho2011thermoelectric} mainly due to their strongly anharmonic lattice that implies a low 
lattice thermal conductivity\cite{delaire2011giant,Li2014phonon,iizumi1975phase,PhysRevB.95.144101,ribeiro2017anharmonic}. PbTe is an appropriate example of the potential technological relevance of thermoelectric 
monochalcogenides: it shows a high $ZT$ in the $600-800$ K temperature range\cite{ravich2013semiconducting}, as high as 2.2 when nanostructured\cite{Hsu818}, and  has been successfully applied in 
spacecrafts\cite{rowe2018thermoelectrics}. In the last years SnSe has attracted a great deal of attention since it was measured to be the most efficient intrinsic thermoelectric material\cite{zhao2014ultralow}. Its figure of merit 
soars to $2.6$ after a structural phase transition\cite{zhao2014ultralow,adouby1998structure,chattopadhyay1986neutron,von1981high} at around $800$ K from the low-symmetry $Pnma$ phase to the high-symmetry $Cmcm$. In the 
high-symmetry phase the electronic band gap is reduced without affecting its ultralow thermal conductivity, providing the record $ZT$. A recent theoretical work shows that the phase transition\cite{aseginolaza2019phonon} is second 
order and non-perturbative anharmonicity is very important to get a thermal conductivity in agreement with experiments. 

SnS is isoelectronic to SnSe and shows very similar electronic and vibrational properties\cite{guo2015first,chattopadhyay1986neutron,von1981high} at low temperatures. Experimentally it also shows a phase 
transition\cite{chattopadhyay1986neutron,von1981high} from the $Pnma$ to the $Cmcm$ structure and a very low thermal conductivity in the former\cite{he2018remarkable,tan2014thermoelectrics} phase. Therefore, it is expected to be a 
very efficient thermoelectric material in the high temperature phase, which together with the fact that S is a much more earth abundant element than Se, makes it a very interesting candidate for technological 
applications. Actually, in Refs. \onlinecite{he2018remarkable,tan2014thermoelectrics} it is shown how the $ZT$ of undoped SnS increases very fast before the phase transition as in the case of SnSe. However, as far as we are 
aware, there are no experimental transport measurements for the high temperature phase of SnS. First-principles calculations of its thermoelectric properties are also absent in the literature, hindered by the unstable 
modes obtained within the harmonic approximation as in the case of SnSe\cite{skelton2016anharmonicity,dewandre2016two}. 

In this work, by performing {\it ab initio} calculations we propose that $Cmcm$ SnS is expected to be a very efficient intrinsic thermoelectric material, as good as SnSe in this phase. We show that the $P_{F}$ of 
SnSe and SnS are expected to be very similar in this phase, as long as the electronic relaxation time is similar in both materials. By including anharmonicity in the phonon calculation at a non-perturbative level within the Stochastic 
Self-Consistent Harmonic Approximation\cite{errea2014anharmonic,bianco2017second,monacelli2018pressure} (SSCHA), we show that the phonon spectrum of SnS suffers a strong anharmonic renormalization. The phase transition is driven by 
the collapse of a zone-border phonon. Anharmonicity is so large that the spectral function of some vibrational modes deviates from the Lorentzian-like shape as it happens in other 
monochalcogenides\cite{Li2014phonon,ribeiro2017anharmonic}. Finally, we calculate the lattice thermal conductivity of $Cmcm$ SnS obtaining ultralow values below $\approx 1.0$ Wm$^{-1}$K$^{-1}$. Non-perturbative anharmonic 
corrections to the third-order force-constants are important in its calculation as it happens in SnSe\cite{aseginolaza2019phonon}. There is a clear anisotropy between in-plane and out-of-plane thermal conductivities. The similarity 
of the power factors and the lattice thermal conductivities of SnSe and SnS suggest that the latter may be an earth abundant efficient thermoelectric material and motivate more experimental effort to measure its $ZT$ in the 
high-temperature phase.

This article is organized as follows. In section \ref{theory} we briefly review the theoretical background for the calculation of anharmonic phonons, thermal conductivity, and electronic transport properties. In 
section \ref{abinitio} we specify the computational details. In section \ref{results} we discuss the results of our work. Conclusions are given in section \ref{conclusions}.

\section{Theoretical Background}
\label{theory}

\subsection{Electronic transport properties}

Within the semiclassical Boltzmann transport theory\cite{PhysRevB.68.125210} the electrical conductivity and the Seebeck coefficient can be calculated respectively as
\begin{eqnarray}
 \sigma(T,\mu) & = & e^2 \int_{-\infty}^{\infty} d\varepsilon \left[-\frac{\partial f(T,\mu,\varepsilon)}{\partial\epsilon}\right] \Sigma(\varepsilon) 
 \label{eq-sigma}\\
 S(T,\mu) & = & \frac{e}{T\sigma(T,\mu)} \int_{-\infty}^{\infty} d\varepsilon \left[-\frac{\partial f(T,\mu,\varepsilon)}{\partial\epsilon}\right] \Sigma(\varepsilon) (\varepsilon - \mu), \nonumber \\
\label{eq-s}
\end{eqnarray} 
where $e$ is the electron charge,  $\mu$ the chemical potential, $ f(T,\mu,\varepsilon)$ the Fermi-Dirac distribution function, and $\Sigma(\varepsilon)$ the transport distribution function. The latter is defined as
\begin{equation}
\Sigma(\varepsilon) = \frac{1}{\Omega N_{\mathbf{k}}} \sum_{n\mathbf{k}} \tau^e_{n\mathbf{k}}  |\mathbf{v}_{n\mathbf{k}}|^{2} \delta( \varepsilon-\varepsilon_{n\mathbf{k}} ), 
\label{eq-tdf}
\end{equation}  
where $\Omega$ is the unit cell volume, $N_{\mathbf{k}}$ the number of $\mathbf{k}$ points in the sum, and $\varepsilon_{n\mathbf{k}} $, $\mathbf{v}_{n\mathbf{k}} $ and  $\tau^e_{n\mathbf{k}}$ are, respectively, the energy, Fermi 
velocity and relaxation time of the electronic state with band index $n$ and crystal momentum $\mathbf{k}$. Our goal here is to compare the power factors $P_{F}(T,\mu)=\sigma(T,\mu) S^2(T,\mu)$ of SnSe and SnS coming from 
their different band structure without explicitly calculating the electronic relaxation times. We thus assume that $\tau^e_{n\mathbf{k}}=\tau^e$ is just the same constant for both compounds. In these conditions it is easy to see 
from Eqs. \eqref{eq-sigma}-\eqref{eq-tdf} that the power factor is proportional to $\tau^e$.  In the following we will limit ourselves to the analysis of $P_{F}(T,\mu)/\tau^e$, which only depends on the band 
structure of the compounds.  

\subsection{Free energy of strongly anharmonic crystals}

We study the vibrational properties of SnS within the Born-Oppenheimer (BO) approximation, in which the Hamiltonian $H$ that determines the dynamics of the ions consists of the ionic kinetic energy and the BO 
potential $V(\mathbf{R})$. $\mathbf{R}$ denotes $R^{\alpha s}(\mathbf{l})$ in component free notation, which specifies the atomic configuration of the crystal. $\alpha$ is a Cartesian direction, $s$ labels an atom within the unit 
cell, and $\mathbf{l}$ a lattice vector. From now on, we will use a single composite index $a=(\alpha,s,\mathbf{l})$ and bold letters to indicate quantities in component-free notation. We will keep this composite index for Fourier 
transformed components adding a bar, $\bar{a}$, to distinguish that in this case $\bar{a}$ just denotes a Cartesian index and an atom in the unit cell.

As it will be shown below and as it happens in $Cmcm$ SnSe\cite{aseginolaza2019phonon,skelton2016anharmonicity,dewandre2016two}, the harmonic approximation collapses for $Cmcm$ SnS. Truncating the Taylor expansion 
of $V(\mathbf{R})$ for this phase at second order and diagonalizing the resulting harmonic force-constants $\boldsymbol{\phi}$ large imaginary frequencies are obtained. This makes the calculation of any thermodynamic and transport 
property involving phonons impossible at the harmonic level. We overcome this problem by solving the ionic Hamiltonian within the SSCHA, a variational method that includes anharmonic effects at a non-perturbative level in the 
calculation of the vibrational free energy\cite{errea2014anharmonic,bianco2017second,monacelli2018pressure}.

The SSCHA performs a variational minimization of the free energy with respect to a trial density matrix $\rho_{\mathcal{H}}$ that solves an auxiliary harmonic Hamiltonian
\begin{equation}
 \mathcal{H}=\sum_{a}\frac{P_{a}^{2}}{2M_{a}}+\frac{1}{2}\sum_{ab}(\mathbf{R}-\boldsymbol{\mathcal{R}})_{a}\Phi_{ab}(\mathbf{R}-\boldsymbol{\mathcal{R}})_{b},
 \label{eq-trialH}
\end{equation}
where $P$ is the kinetic energy and $M_a$ the atomic mass of atom $a$. 
The variational parameters in the minimization are the $\mathbf{\Phi}$ force-constants, which should not be confused with the harmonic force-constants $\boldsymbol{\phi}$,  and the $\boldsymbol{\mathcal{R}}$ 
positions. The $\boldsymbol{\mathcal{R}}$ positions are referred as the \emph{centroid} positions, i.e., the most probable atomic positions. The $\mathbf{\Phi}$ force-constants are related to the broadening of the ionic wave 
functions around the centroid positions. At the minimum, the SSCHA yields a free energy $F$ that takes into account anharmonicity without approximating the BO potential. The minimization can be performed by calculating atomic 
forces and energies in stochastic atomic configurations in supercells using importance sampling and reweighting techniques\cite{errea2014anharmonic,bianco2017second,monacelli2018pressure}. The supercell atomic configurations 
are created  according to the probability distribution function related to $\rho_{\mathcal{H}}$. Since the BO energy landscape is sampled stochastically, the SSCHA method does not use any fit or approximation on 
the $V(\boldsymbol{R})$. It is, therefore, unbiased by the starting point. 

\subsection{Free energy Hessian and second-order phase transition}

In a displacive second-order phase transition, at high temperature the free energy $F$ has a minimum in a high-symmetry configuration ($\boldsymbol{\mathcal{R}}_{hs}$), but, on lowering the 
temperature, $\boldsymbol{\mathcal{R}}_{hs}$ becomes a saddle point at the transition temperature $T_{c}$. Therefore, the free energy Hessian evaluated at $\boldsymbol{\mathcal{R}}_{hs}$, 
$\partial^{2}F/\partial\boldsymbol{\mathcal{R}}\partial\boldsymbol{\mathcal{R}}|_{\boldsymbol{\mathcal{R}}_{hs}}$, at high temperature is positive definite but it develops one or multiple negative eigendirections at $T_{c}$. The 
SSCHA free energy Hessian can be computed by using the analytic formula\cite{bianco2017second} 
\begin{equation}\label{formula}
 \frac{\partial^{2}F}{\partial\boldsymbol{\mathcal{R}}\partial\boldsymbol{\mathcal{R}}}=\mathbf{\Phi} + \overset{(3)}{\mathbf{\Phi}}\mathbf{
 \Lambda}(0)[\mathbf{1}-\overset{(4)}{\mathbf{\Phi}}\mathbf{\Lambda}(0)]^{-1}\overset{(3)}{\mathbf{\Phi}},
\end{equation}
with
\begin{equation}\label{fcnp}
 \overset{(n)}{\mathbf{\Phi}}=\left\langle\frac{\partial^{n}V}{\partial\mathbf{R}^{n}}\right\rangle_{\rho_{\mathcal{H}}}. 
\end{equation}
Here $\left\langle\right\rangle_{\rho_{\mathcal{H}}}$ denotes the quantum statistical average taken with the density matrix $\rho_{\mathcal{H}}$. All these averages are evaluated here stochastically as described in 
Ref. \onlinecite{bianco2017second}. The $\overset{(n)}{\mathbf{\Phi}}$ non-perturbative $n$-th order force-constants should not be confused with the $n$-th order perturbative 
force-constants $\overset{(n)}{\boldsymbol{\phi}}$, which are calculated as derivatives of the BO potential at a reference position $0$ and not as quantum statistical averages:
 \begin{equation}\label{fcp}
 \overset{(n)}{\boldsymbol{\phi}}=\left[\frac{\partial^{n}V}{\partial\mathbf{R}^{n}}\right]_0. 
\end{equation}
In Eq. \eqref{formula} the value $z=0$ of the fourth-order tensor $\mathbf{\Lambda}(z)$ is used. For a generic complex number $z$ it is defined, in components, by
\begin{multline}
 \Lambda^{abcd}(z)=-\frac{1}{2}\sum_{\mu\nu}\tilde{F}(z,\tilde{\Omega}_{\mu},\tilde{\Omega}_{\nu})\times \\ \sqrt{\frac{\hbar}{2M_{a}\tilde{
 \Omega}_{\mu}}}e_{\mu}^{a}\sqrt{\frac{\hbar}{2M_{b}\tilde{\Omega}_{\nu}}}e_{\nu}^{b}\sqrt{\frac{\hbar}{2M_{c}\tilde{\Omega}_{\mu}}}e_{\mu}^{
 c}\sqrt{\frac{\hbar}{2M_{d}\tilde{\Omega}_{\nu}}}e_{\nu}^{d},
\label{eq-lambda}
\end{multline}
with  $\tilde{\Omega}_{\mu}^{2}$ and $e_{\mu}^{a}$ the eigenvalues and corresponding eigenvectors of
\begin{equation}
D_{ab}^{(S)}=\Phi_{ab}/\sqrt{M_{a}M_{b}},
\label{eq-ds}
\end{equation}
respectively. In Eq. \eqref{eq-lambda}  
\begin{multline}
 \tilde{F}(z,\tilde{\Omega}_{\mu},\tilde{\Omega}_{\nu})=\frac{2}{\hbar}\left[\frac{(\tilde{\Omega}_{\mu}+\tilde{\Omega}_{\nu})[1+n_{B}(\tilde{
 \Omega}_{\mu})+n_{B}(\tilde{\Omega}_{\nu})]}{(\tilde{\Omega}_{\mu}+\tilde{\Omega}_{\nu})^{2}-z^{2}} - \right. \\ \left. \frac{(\tilde{\Omega}_{\mu}-\tilde{
 \Omega}_{\nu})[n_{B}(\tilde{\Omega}_{\mu})-n_{B}(\tilde{\Omega}_{\nu})]}{(\tilde{\Omega}_{\mu}-\tilde{\Omega}_{\nu})^{2}-z^{2}}\vphantom{\int_1^2}\right],
\end{multline}
where $n_{B}(\omega)=1/(e^{\beta\hbar\omega}-1)$ is the bosonic occupation number. Evaluating through Eq. \eqref{formula} the free energy Hessian at $\boldsymbol{\mathcal{R}}_{hs}$ and studying its spectrum as a function of 
temperature, we can predict the occurrence of a displacive phase transition and estimate $T_{c}$. This technique has been successful to study phase-transition temperatures in high-pressure hydrides, monochalcogenides, and transition 
metal dichalcogenides undergoing charge-density wave transitions\cite{ribeiro2017anharmonic,PhysRevB.97.214101,doi:10.1021/acs.nanolett.9b00504}. 

\subsection{Dynamical properties of solids and phonon frequencies}

As shown in Ref. \onlinecite{bianco2017second}, even if the SSCHA is a ground-state theory, it is possible to formulate a valid ansatz in order to calculate dynamical properties of crystals such as phonon spectral functions. The 
one-phonon Green function $\mathbf{G}(z)$ for the variable $\sqrt{M_{a}}(R_{a}-\mathcal{R}_{a})$ can be calculated as 
\begin{equation} \label{green}
 \mathbf{G}^{-1}(z)=z^{2}\mathbf{1}-\mathbf{M}^{-\frac{1}{2}}\mathbf{\Phi}\mathbf{M}^{-\frac{1}{2}}- \mathbf{\Pi}(z).
\end{equation}
With this definition, in the static limit the Green function becomes the dynamical matrix obtained with the free energy Hessian: $\mathbf{G}^{-1}(0)=-\mathbf{D}^{(F)}$, with 
\begin{equation}
D_{ab}^{(F)}=\frac{1}{\sqrt{M_{a}M_{b}}}\frac{\partial^{2}F}{\partial\mathbf{\mathcal{R}}_{a}
\partial\mathbf{\mathcal{R}}_{b}}.
\end{equation} 
We will label with $\omega_{\mu}$ the eigenvalues of $\mathbf{D}^{(F)}$. The SSCHA self-energy is given by
\begin{equation}
 \Pi(z)=\mathbf{M}^{-\frac{1}{2}}\overset{(3)}{\mathbf{\Phi}}\mathbf{\Lambda}(z)[\mathbf{1}-\overset{(4)}{\mathbf{\Phi}}\mathbf{\Lambda}(z)]^{-1}
 \overset{(3)}{\mathbf{\Phi}}\mathbf{M}^{-\frac{1}{2}},
\end{equation}
where $M_{ab}=\delta_{ab}M_{a}$ is the mass matrix. 
We have explicitly verified that neglecting $\overset{(4)}{\boldsymbol{\Phi}}$ in Eq. \eqref{formula} has a completely negligible impact on the eigenvalues of $\mathbf{D}^{(F)}$. We consistently  
neglect $\overset{(4)}{\boldsymbol{\Phi}}$ in Eq. \eqref{green} as well. This reduces the SSCHA self energy to the so-called bubble self energy, namely
\begin{equation}
 \mathbf{\Pi}(z)\approx\mathbf{\Pi}^{(B)}(z)=\mathbf{M}^{-\frac{1}{2}}\overset{(3)}{\mathbf{\Phi}}\mathbf{\Lambda}(z)\overset{(3)}{\mathbf{
 \Phi}}\mathbf{M}^{-\frac{1}{2}}.
\end{equation}

The cross section in an inelastic, e.g. neutron, experiment is proportional to the spectral function $\sigma(\omega)=-\omega \mathrm Tr Im\mathbf{G}(\omega+i0^{+})/\pi$\cite{0034-4885-31-1-303}. Its peaks signal the presence of 
collective vibrational excitations (phonons) having certain energies and linewidth. In order to recognize the contribution of each phonon mode to this spectral function, we first take advantage of the lattice periodicity and 
Fourier transform the spectral function and the self energy, and second we neglect the mixing between phonon modes and assume that $\mathbf{\Pi}(z)$ is diagonal in the basis of the eigenvectors:
\begin{equation}
\Pi_{\mu}(\mathbf{q},\omega)=\sum_{\bar{a}\bar{b}}e_{\mu}^{\bar{a}}(-\mathbf{q})\Pi_{\bar{a}\bar{b}}(\mathbf{q},\omega+i0^{+})e_{\mu}^{\bar{b}}
 (\mathbf{q}).
 \label{eq-pi-mode}
\end{equation}
The cross section is then given by
\begin{multline}\label{spectral}
 \sigma(\mathbf{q},\omega)=\\ \frac{1}{\pi}\sum_{\mu}\frac{-\omega Im\Pi_{\mu}(\mathbf{q},\omega)}{(\omega^{2}-\tilde{\Omega}_{\mu}^{2}(\mathbf{q})-
 Re\Pi_{\mu}(\mathbf{q},\omega))^{2}+(Im\Pi_{\mu}(\mathbf{q},\omega))^{2}}.
\end{multline}
In Eqs. \eqref{eq-pi-mode} and \eqref{spectral} $\tilde{\Omega}_{\mu}^{2}(\mathbf{q})$ and $e_{\mu}^{\bar{a}}(\mathbf{q})$ are, respectively, the eigenvalues and eigenvectors of  $\mathbf{D}^{(S)}(\mathbf{q})$, the Fourier 
transform of Eq. \eqref{eq-ds}.

The cross section calculated as in Eq. \eqref{spectral} does not have any given lineshape. However, when  $\Pi_{\mu}(\mathbf{q},\omega)$ is small  compared to $\tilde{\Omega}_{\mu}^{2}(\mathbf{q})$, it is justified to 
approximate  $\Pi_{\mu}(\mathbf{q},\omega) \sim \Pi_{\mu}(\mathbf{q},\tilde{\Omega}_{\mu}(\mathbf{q}))$, which turns $ \sigma(\mathbf{q},\omega)$ into a sum of Lorentzian functions. In this Lorentzian approximation the peaks appear 
at the $\Omega_{\mu}(\mathbf{q})$ phonon frequencies, with
\begin{equation}\label{lorentzian}
\Omega_{\mu}^{2}(\mathbf{q})=\tilde{\Omega}_{\mu}^{2}(\mathbf{q})+Re\Pi_{\mu}(\mathbf{q},\tilde{\Omega}_{\mu}(\mathbf{q})), 
\end{equation}
and the linewidths are proportional to $Im[\Pi_{\mu}(\mathbf{q},\tilde{\Omega}_{\mu}(\mathbf{q}))]$.

\subsection{Thermal conductivity}

We calculate the lattice thermal conductivity within the single mode relaxation time approximation (SMA)\cite{doi:10.1063/1.1391230} making use of the eigenvalues and eigenvectors of $\mathbf{D}^{(S)}(\mathbf{q})$ (as it will be 
shown below it is not possible at the harmonic level due to the instabilities obtained) as well as the non-preturbative third-order force-constants $\overset{(3)}{\mathbf{\Phi}}$. In the SMA the lattice thermal conductivity is 
written as follows\cite{paulatto2013anharmonic}:
\begin{multline} \label{smatk}
\kappa_{l}^{\alpha\beta}=\frac{\hbar^{2}}{\Omega N_{\mathbf{q}}k_{B}T^{2}}\times \\ \sum_{\mathbf{q}\mu}c_{\mu}^{\alpha}(\mathbf{q})c_{\mu}^{\beta}(\mathbf{q})
\tilde{\Omega}_{\mu}^{2}(\mathbf{q})n_{B}(\tilde{\Omega}_{\mu}(\mathbf{q}))\left[n_{B}(\tilde{\Omega}_{\mu}(\mathbf{q}))+1\right]\tau_{\mu}(\mathbf{q}),
\end{multline}
where, for the phonon mode $\mu$ with momentum $\mathbf{q}$, $c_{\mu}^{\alpha}(\mathbf{q})$ is the Cartesian component $\alpha$ of its lattice group velocity and $\tau_{\mu}(\mathbf{q})$ its lifetime. $N_{\mathbf{q}}$ is the number 
of $\mathbf{q}$ points used in the sum. The Bose-Einstein occupation of each mode is given by the Boltzmann Transport Equation (BTE) and the inverse phonon lifetime 
(with $\gamma_{\mu}(\mathbf{q})$ the half width at half maximum) is calculated 
as\cite{paulatto2013anharmonic}
\begin{multline} \label{liw}
\frac{1}{\tau_{\mu}(\mathbf{q})}=2\gamma_{\mu}(\mathbf{q})=\frac{\pi}{\hbar^{2}N_{\mathbf{q}}} \sum_{\mathbf{q}'\nu\eta}|\overset{(3)}{\Phi}_{\mu\nu\eta}(
\mathbf{q},\mathbf{q}',\mathbf{q}'')|^{2} \\ 
\times [ 
(1+n_{B}(\tilde{\Omega}_{\nu}(\mathbf{q}'))+n_{B}(\tilde{\Omega}_{\eta}(\mathbf{q}''))) \delta(\tilde{\Omega}_{\mu}(\mathbf{q})-\tilde{\Omega}_{\nu}(\mathbf{q}')-\tilde{\Omega}_{\eta}(\mathbf{q}''))
 \\ +2 (n_{B}(\tilde{\Omega}_{\nu}(\mathbf{q}'))-n_{B}(\tilde{\Omega}_{\eta}(\mathbf{q}'')))  \delta ( \tilde{\Omega}_{\mu}(\mathbf{q})+\tilde{\Omega}_{\nu}(\mathbf{q}')-\tilde{\Omega}_{\eta}(\mathbf{q}'') ) 
],
\end{multline}
with $\mathbf{q}+\mathbf{q}'+\mathbf{q}''=\mathbf{G}$, $\mathbf{G}$ being a reciprocal lattice vector.
Here $\overset{(3)}{\Phi}_{\mu\nu\eta}(\mathbf{q},\mathbf{q}',\mathbf{q}'')$ is the third order force-constants matrix written in the space of the normal modes: 
\begin{multline}
\overset{(3)}{\Phi}_{\mu\nu\eta}(\mathbf{q},\mathbf{q}',\mathbf{q}'') = \sum_{\bar{a}\bar{b}\bar{c}} \sqrt{\frac{\hbar^3}{8 M_{\bar{a}} M_{\bar{b}} M_{\bar{c}} \tilde{\Omega}_{\mu}(\mathbf{q}) \tilde{\Omega}_{\nu}(\mathbf{q}') 
\tilde{\Omega}_{\eta}(\mathbf{q}'')}} \\ \times e^{\bar{a}}_{\mu}(\mathbf{q}) e^{\bar{b}}_{\nu}(\mathbf{q}') e^{\bar{c}}_{\eta}(\mathbf{q}'') \overset{(3)}{\Phi}_{\bar{a}\bar{b}\bar{c}}(\mathbf{q},\mathbf{q}',\mathbf{q}''),
\label{eq-fcmode}
\end{multline}
where $\overset{(3)}{\Phi}_{\bar{a}\bar{b}\bar{c}}(\mathbf{q},\mathbf{q}',\mathbf{q}'')$ are the Fourier transformed non-perturbative third-order force-constants. We also calculate the thermal conductivity with the perturbative 
third-order force-constants by substituting the non-perturbative $\overset{(3)}{\mathbf{\Phi}}$ by the perturbative $\overset{(3)}{\boldsymbol{\phi}}$ in Eqs. \eqref{liw} and \eqref{eq-fcmode}.

\section{Computational Details}
\label{abinitio}

\begin{figure}[ht]
\includegraphics[width=\linewidth]{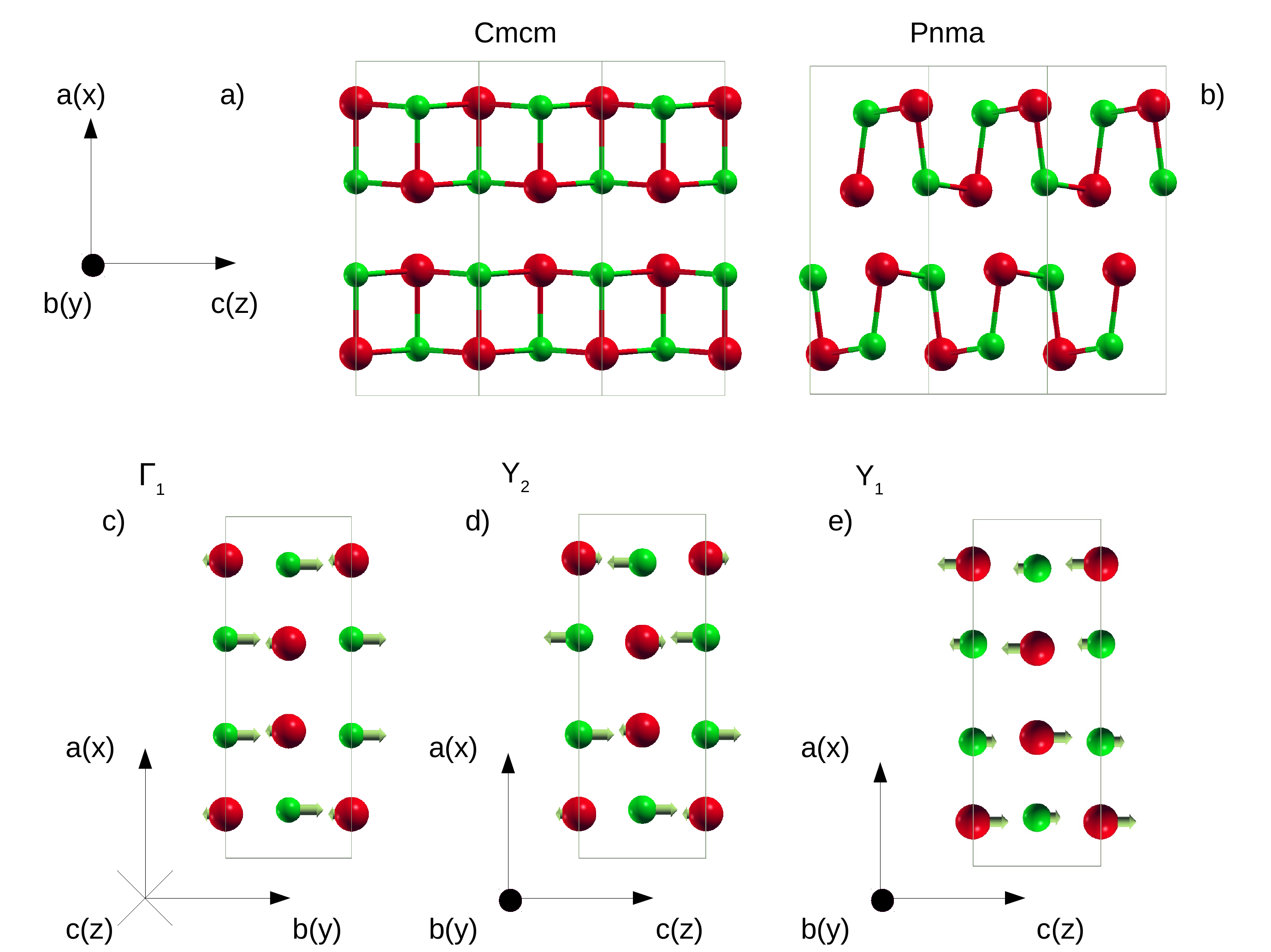}
\caption{XY face of the a) $Cmcm$ and b) $Pnma$ structures.  Atomic displacements of modes c) $\Gamma_{1}$, d) $Y_{2}$ and e) $Y_{1}$. Sn atoms are red and S green.} 
\label{structure}
\end{figure}

We calculate the electronic bands using {\it ab initio} Density Functional Theory (DFT) calculations within the local density approximation (LDA)\cite{perdew1981self} and the generalized gradient approximation in the Perdew Burke 
Ernzerhof (PBE) parametrization\cite{perdew1996generalized} as implemented in the {\sc Quantum ESPRESSO}\cite{giannozzi2009quantum,Giannozzi_2017} software package. Harmonic phonons and perturbative third-order 
force-constants $\overset{(3)}{\boldsymbol{\phi}}$ are calculated  using Density Functional Perturbation Theory\cite{baroni2001phonons,paulatto2013anharmonic}. We use projector augmented wave\cite{blochl1994projector} (PAW) 
pseudopotentials that include $5s^{2}$ $5p^{2}$ $4d^{10}$ electrons in the case of Sn and $3s^{2}$ $3p^{4}$ in the case of S or Se. For the perturbative third-order force-constants we use norm-conserving pseudopotentials which were 
shown\cite{aseginolaza2019phonon} to provide very similar third-order force-constants compared to the PAW result. A $16\times16\times16$ sampling of the first Brillouin zone of the primitive cell and an energy 
cutoff of $70$ Ry are employed for the DFT self-consistent calculation. For the electronic transport calculations we use the Boltztrap software package\cite{madsen2006boltztrap}. For the sum in Eq. \ref{eq-tdf} we perform a non 
self-consistent DFT calculation in a $30\times30\times30$ sampling of the first Brillouin zone. We use experimental lattice parameters at the transition temperature as we got better agreement with experiments for SnSe in a previous 
work\cite{aseginolaza2019phonon}. The experimental lattice parameters taken from Refs.\cite{adouby1998structure,chattopadhyay1986neutron} are $a=22.13$ a$_{0}$, $b=8.13$ a$_{0}$, $c=8.13$ a$_{0}$ for SnSe 
and $a=21.69$ a$_{0}$, $b=7.84$ a$_{0}$, $c=7.84$ a$_{0}$ (a$_{0}$ is the Bohr length) for SnS. The structures of the high temperature $Cmcm$ and low temperature $Pnma$ phases are shown in Figure \ref{structure}.

Anharmonic phonons and non-perturbative third-order force-constants are calculated within the SSCHA using a $2\times2\times2$ supercell. For the SSCHA calculation we use forces calculated within DFT. Once we get the anharmonic 
force-constants, we substract the harmonic ones and interpolate the difference to a $6\times6\times6$ grid. Then, we add this interpolated difference to the harmonic dynamical matrices in a $6\times6\times6$ grid, which yields 
anharmonic force-constants in a fine grid. By Fourier interpolating the latter we can calculate phonon frequencies at any point in the Brillouin zone. We impose the acoustic sum rule to the third-order force-constants with an 
iterative method prior to their Fourier interpolation\cite{paulatto2013anharmonic,aseginolaza2019phonon}. The lattice thermal conductivity is calculated with Eq. \eqref{smatk} using a $10\times10\times10$ grid. For the 
calculation of the phonon linewidths we use a $20\times20\times20$ mesh in Eq. \eqref{liw} with a Gaussian smearing of 1 cm$^{-1}$ for the Dirac deltas. 

\section{Results and Discussion}
\label{results}

\subsection{Electronic transport}

\begin{figure*}[ht]
\includegraphics[width=0.45\linewidth]{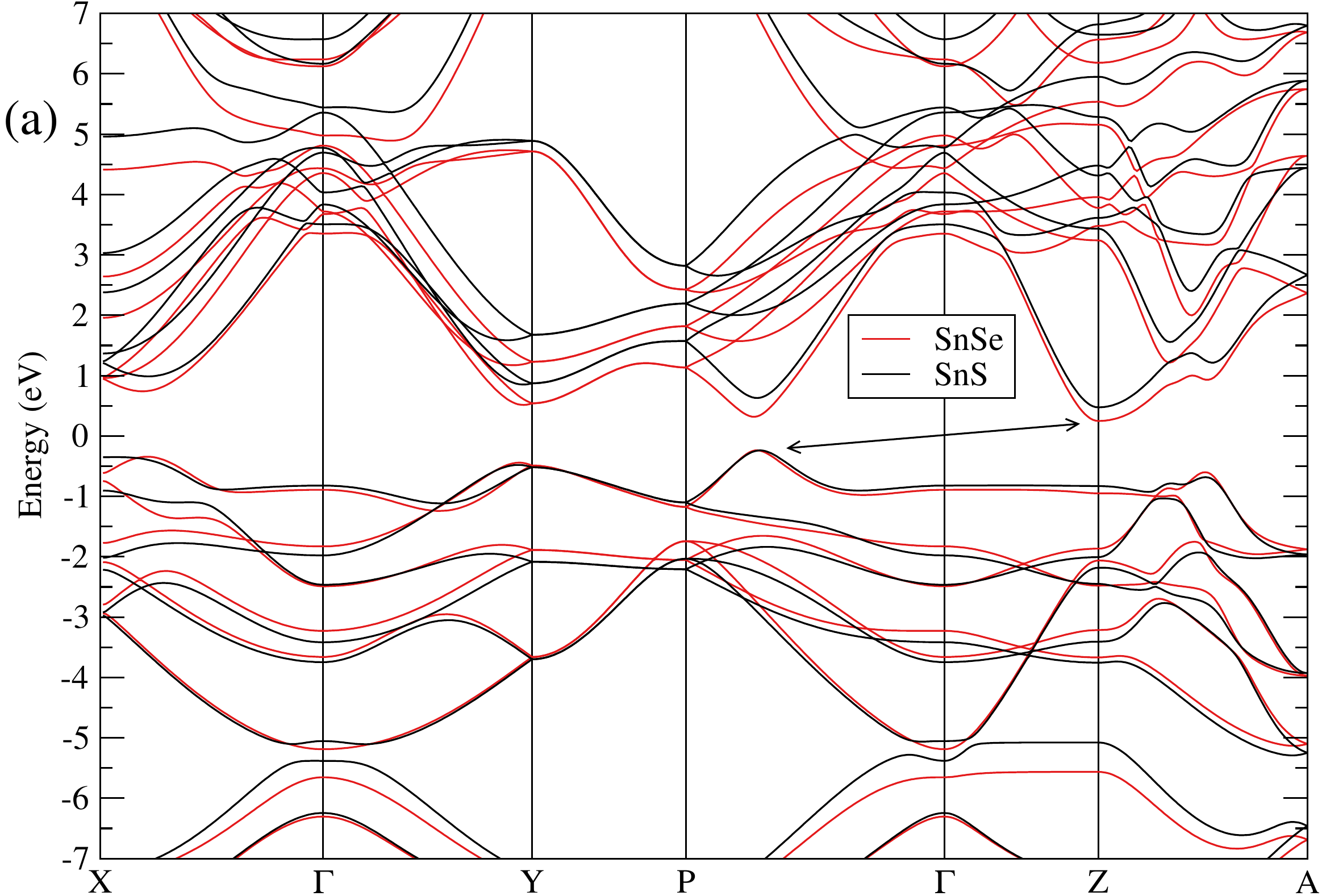}
\includegraphics[width=0.43\linewidth]{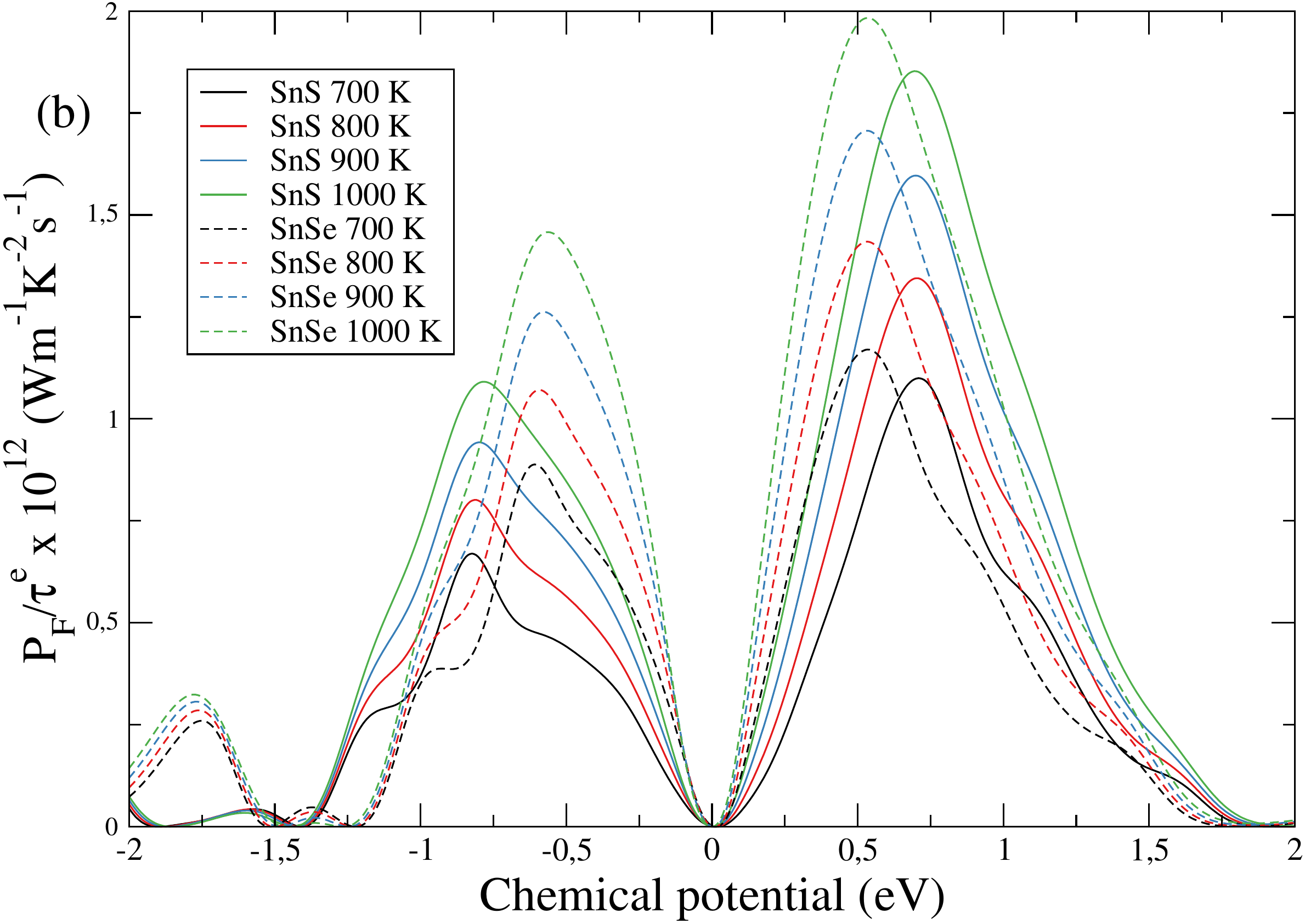}
\caption{(a) Electronic band structure of $Cmcm$ SnS and SnSe using experimental lattice parameters. (b) $PF/\tau^{e}$ of $Cmcm$ SnS and SnSe for different temperatures as a function of the chemical potential. The $0$ value 
corresponds to the middle of the gap in both figures.}
\label{bands}
\end{figure*}

Figure \ref{bands} (a) shows the electronic band structures of SnS and SnSe in the high symmetry phase. 
It shows that the electronic properties of these materials are very similar because their electronic band structures are basically the same as expected for isoelectronic compounds with the same atomic structure. The major 
difference is that the indirect (the conduction and valence bands that constitute the gap are denoted with an arrow in Figure \ref{bands} (a)) energy gap ($0.45$ eV for SnSe and $0.7$ eV for SnS) is bigger in the case of SnS, in 
agreement with experiments\cite{vidal2012band,zhao2014ultralow} and previous calculations\cite{guo2015first}. As expected, the calculated electronic gaps within LDA underestimate the experimental values ($0.86-0.948$ eV for SnSe 
and $0.9-1.142$ eV for SnS). Using these band structures we have calculated the Seebeck coefficient, which within the approximation of a constant electronic relaxation time it is independent of it, and the electrical conductivity 
over the electronic relaxation time $\sigma/\tau^{e}$. The Seebeck coefficient is very similar for both materials, but $\sigma/\tau^{e}$ is slightly larger in the case of SnSe due to the smaller electronic gap. Using these two 
quantities we have calculated  $P_{F}/\tau^{e}$, shown in Figure \ref{bands} (b). As we can see,  $P_{F}/\tau^{e}$ is very similar for both materials, but slightly higher in the case of SnSe. As we can 
see,  $P_{F}/\tau^{e}$ increases as temperature increases and the difference between SnSe and SnS is less than $5\%$ at $1000$ K. These results make clear that regarding the electronic transport properties these two 
materials are very similar in the high temperature phase provided that the relaxation time for the electrons is similar for both materials, which is expected for isoelectronic and isostructural compounds. 

\subsection{$Pnma$ to $Cmcm$ phase transition}

As it was already pointed out\cite{chattopadhyay1986neutron,aseginolaza2019phonon}, symmetry\cite{Orobengoa:ks5225,Perez-Mato:sh5107} dictates that it is possible to have a second-order phase transition between 
the $Cmcm$ and $Pnma$ phases. The transition is dominated by the distortion pattern associated to a non-degenerate mode ($Y_{1}$) at the zone border $Y$ point. This means\cite{aseginolaza2019phonon} that, in a second-order 
displacive phase transition scenario, the transition temperature $T_{c}$ is defined as $\partial^{2}F/\partial Q^{2}(T=T_{c})=0$ where $Q$ is the order parameter that transforms the system continuously from the $Pnma$ ($Q\ne0$) to 
the $Cmcm$ ($Q=0$) phase. As the distortion is dominated by the $Y_{1}$ phonon,  $\partial^{2}F/\partial Q^{2}(T)$ is proportional to $\omega_{Y_{1}}^{2}(T)$, which we can calculate diagonalizing $\mathbf{D}^{(F)}$. 

\begin{figure}[ht]
\includegraphics[width=\linewidth]{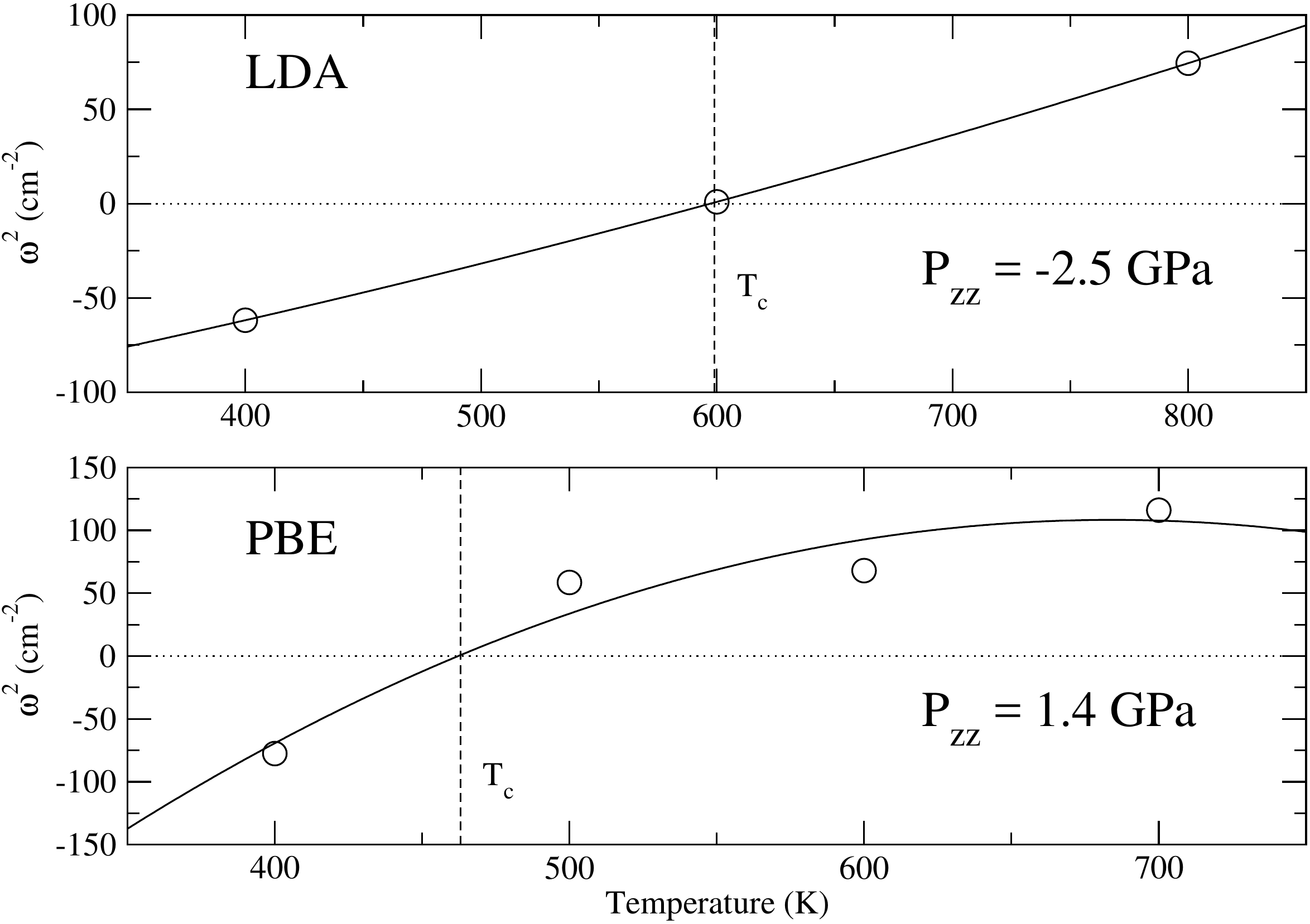}
\caption{$\omega_{Y_{1}}^{2}$ as a function of temperature within LDA and PBE approximations using the experimental lattice parameters (circles). The solid lines correspond to a polynomial fit. We include the pressure component 
$P_{zz}$, which is the pressure in the direction where the atoms move in the transition.  This pressure is calculated including the anharmonic vibrational energy within the SSCHA as discussed in 
Ref. \onlinecite{monacelli2018pressure}.} 
\label{transition}
\end{figure}

Figure \ref{transition} shows $\omega_{Y_{1}}^{2}(T)$ within the LDA and PBE approximations.  
As in the case of SnSe\cite{aseginolaza2019phonon}, the second derivative of the free energy is positive at high temperatures and decreases lowering the temperature. For both approximations, it becomes negative at the critical 
temperature $T_c$, which means that the $Pnma$ phase is not any longer a minimum of the free energy and the structure distorts adopting the $Pnma$ phase. $T_{c}$ strongly depends on the approximation of the exchange-correlation 
functional: it is $600$ K for LDA and $465$ K for PBE. Our LDA calculation agrees better with the experimental value, around $900$ K\cite{chattopadhyay1986neutron}. We associate the discrepancy between LDA and PBE  to the different 
pressures obtained in the transition direction, $P_{zz}$. In fact, as shown in the case of SnSe\cite{aseginolaza2019phonon}, $T_{c}$ depends strongly on the pressure in this $z$ direction. The pressure in 
Figure \ref{transition} includes anharmonic vibrational effects on the energy following the procedure outlined in Ref. \onlinecite{monacelli2018pressure}. For the same lattice parameter LDA displays a much smaller pressure, as 
generally LDA predicts smaller lattice volumes than PBE. The underestimation with respect to experiments may be attributed to the small supercell size used for the SSCHA calculations ($2\times2\times2$). Even if 
experimentally $T_{c}$ is around $100$ K higher in SnS than in SnSe, our LDA calculations give basically the same transition temperature for both materials as $T_{c}=616$ K in SnSe according to our previous 
calculations\cite{aseginolaza2019phonon}. However, within PBE SnSe does show a lower transition temperature since $T_c=299$ K for SnSe K\cite{aseginolaza2019phonon}.

\subsection{Anharmonic phonons}

\begin{figure*}[ht]
\includegraphics[width=0.48\linewidth]{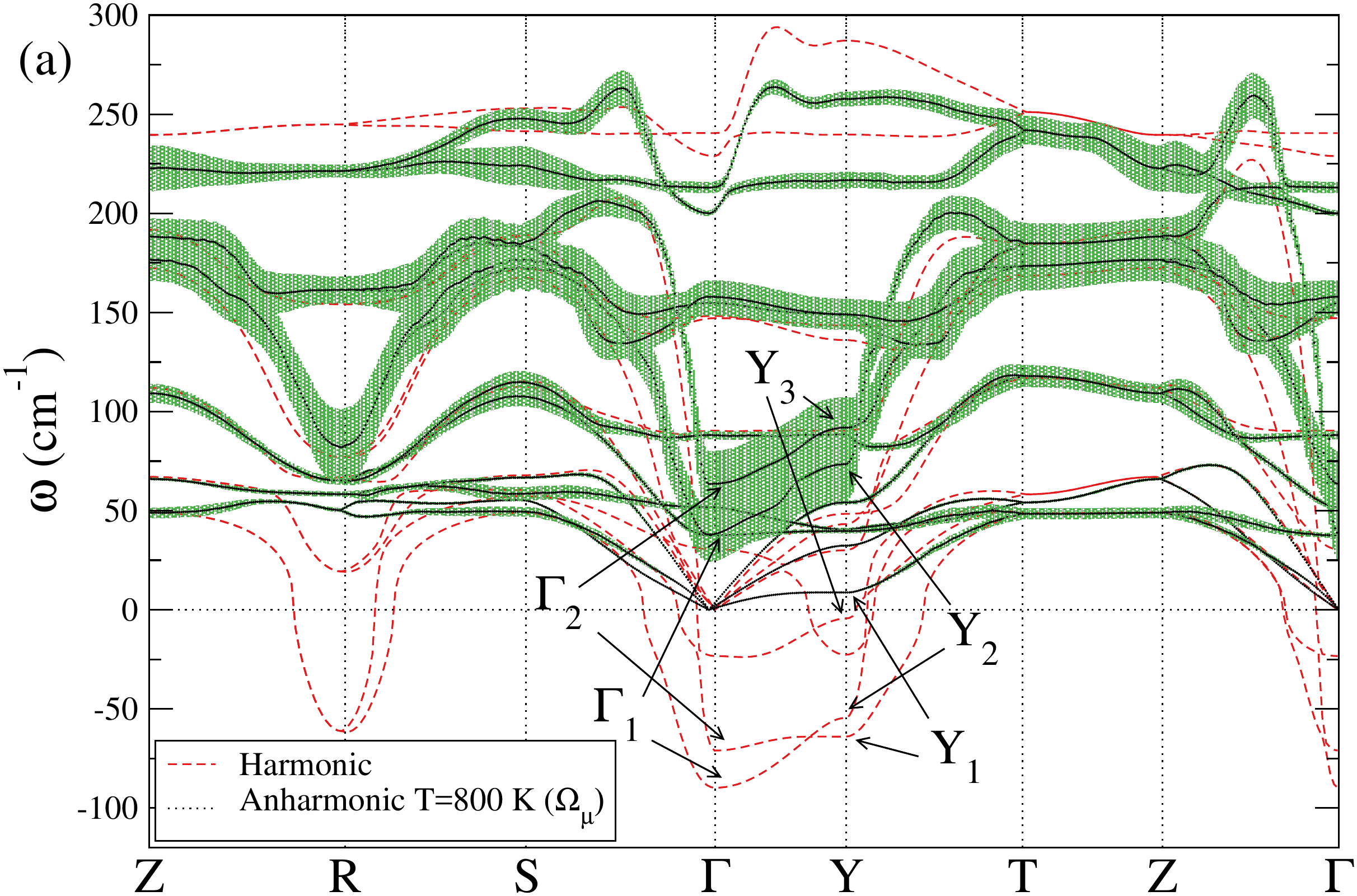}
\includegraphics[width=0.42\linewidth]{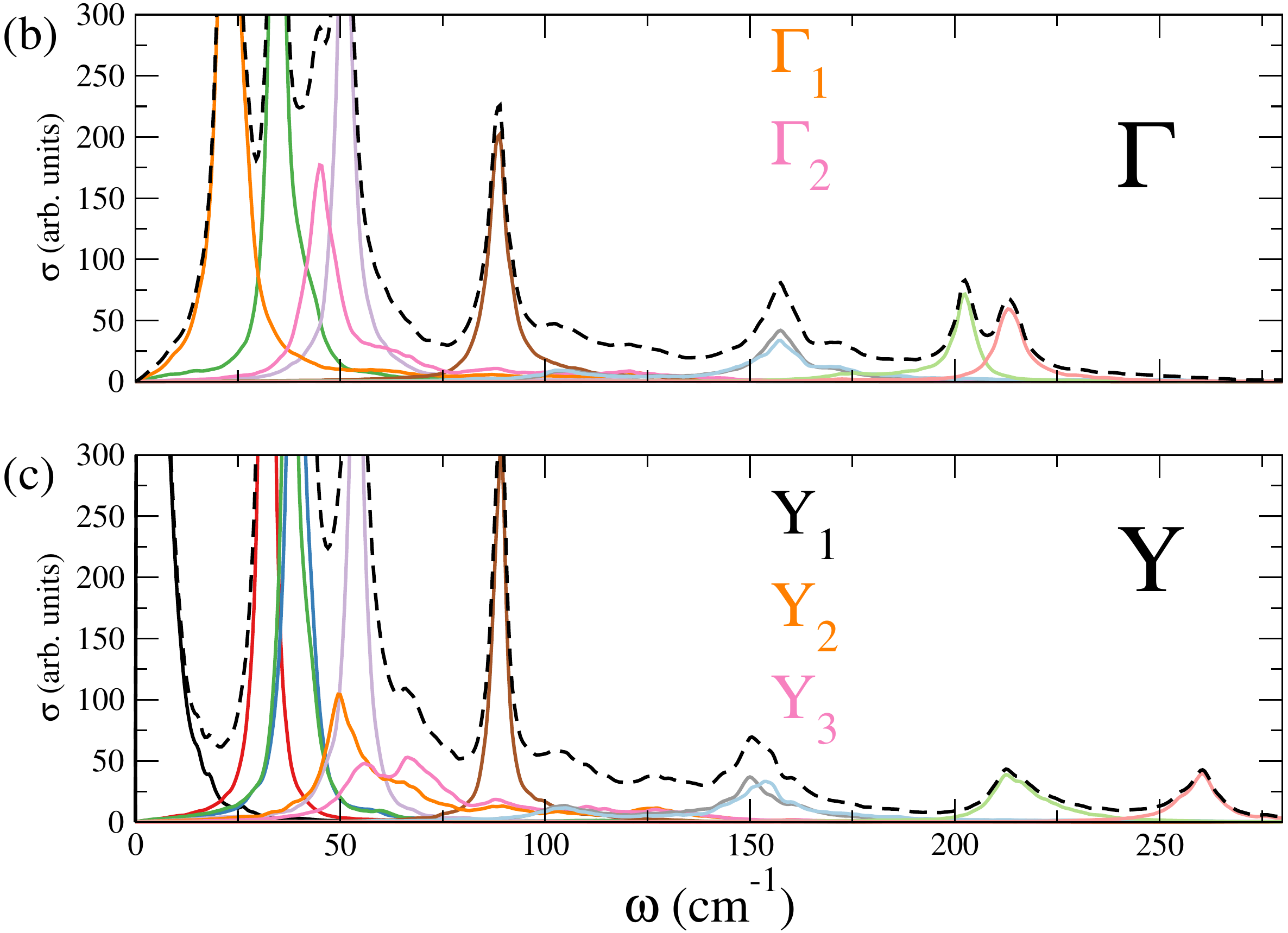}
\caption{(a) Harmonic and anharmonic $[\Omega_{\mu}(\mathbf{q})]$ phonon spectra within the Lorentzian approximation. The length of the bars corresponds to the linewidth (full length of the line is the full width at half 
maximum). The calculations are done within the LDA using $\overset{(3)}{\boldsymbol{\Phi}}$ at $800$ K and $\tilde{\Omega}_{\mu}(\mathbf{q})$ at $800$ K. (b) and (c) $\sigma(\omega)$ spectral functions at the points $\Gamma$ 
and $Y$, respectively, calculated as in Eq. \eqref{spectral}. Solid lines correspond to individual modes and dashed lines are the total spectral functions.} 
\label{lw}
\end{figure*}

Figure \ref{lw} (a) compares the harmonic phonon spectrum with the anharmonic one calculated within the Lorentzian approximation at $800$ K within the LDA. In the anharmonic spectrum shown the phonon energies correspond to 
the $\Omega_{\mu}(\mathbf{q})$ values of Eq. \eqref{lorentzian}. The linewidth obtained in the Lorentzian approximation is also shown. 
The phonon spectrum suffers from a huge anharmonic renormalization. The harmonic spectrum shows broad instabilities, which are stabilized by anharmonicity. The $Y_{1}$ mode is unstable below the transition temperature, but it is 
stabilized after the transition. By having a look at the the phonon linewidths, we can see that two modes at the $\Gamma$ point ($\Gamma_{1}$ and $\Gamma_{2}$) not only suffer a strong anharmonic renormalization, but they also 
have a  large linewidth compared to the rest of the modes in the first Brillouin zone. These modes describe optical in-plane atomic displacements (see Figure \ref{structure}, $\Gamma_{2}$ has the same atomic displacements 
as $\Gamma_{1}$ but in the other in-plane direction), which are the same atomic displacements of $Y_{2}$ and $Y_{3}$ at the point $Y$ with a different periodicity due to the different momentum. The $Y_2$ and $Y_3$ in-plane modes 
also show a very large linewidth. On the contrary, the linewidth of mode $Y_{1}$ is not so large even if it is responsible for the phase transition (see Figure \ref{structure}). 

In strongly anharmonic materials\cite{delaire2011giant,Li2014phonon,ribeiro2017anharmonic,ribeiro2017anharmonic,Li2014phonon,paulatto2015first,PhysRevB.97.214101,aseginolaza2019phonon}, the phonon spectral functions 
$\sigma(\mathbf{q},\omega)$ show broad peaks, shoulders, and satellite peaks that cannot be captured by the simple Lorentzian picture. In Figure \ref{lw} (b) and (c) we show the spectral function keeping the full frequency 
dependence on the self-energy (see Eq. \eqref{spectral}). The calculation is done for the $\Gamma$ and $Y$ points. The great majority of the modes describe a Lorentzian shape. However, the modes with a large linewidth within 
the Lorentzian approximation (see Figure \ref{lw} (a)) are those that clearly deviate from the  Lorentzian profile ($\Gamma_{1}$, $\Gamma_{2}$, $Y_{2}$, $Y_{3}$). This non-Lorentzian shape makes clear that these modes are strongly 
anharmonic and the frequency dependence of the self-energy is crucial to account for their spectral function. In this case, as we can see in in Figures \ref{lw} (b) and (c), the non-Lorentzian shapes of the strongly anharmonic modes 
do not create appreciable shoulders or satellite peak in the total spectral function, however, their contribution is far from trivial.

\subsection{Lattice thermal transport}

\begin{figure}[ht]
\includegraphics[width=\linewidth]{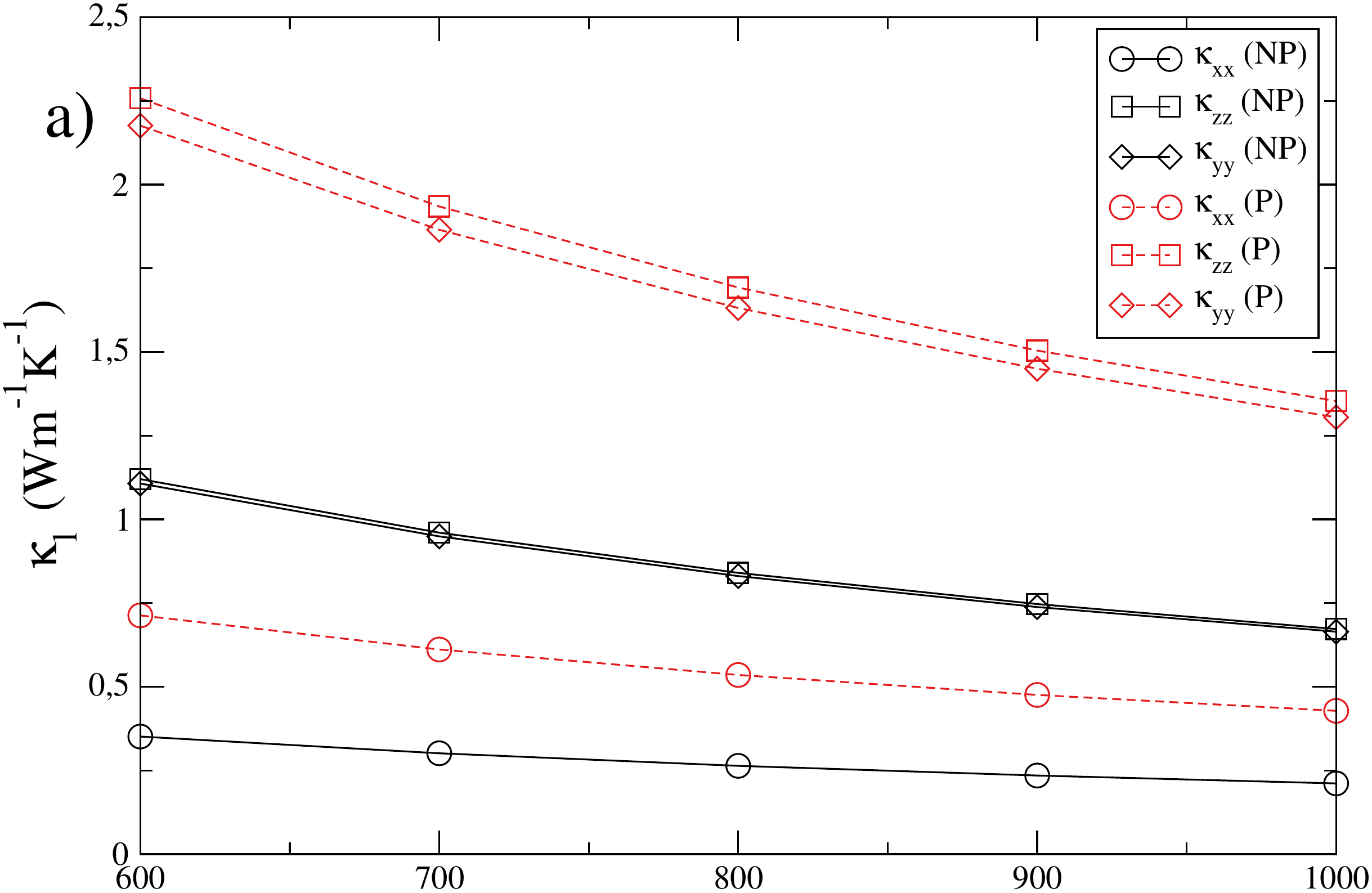}
\includegraphics[width=\linewidth]{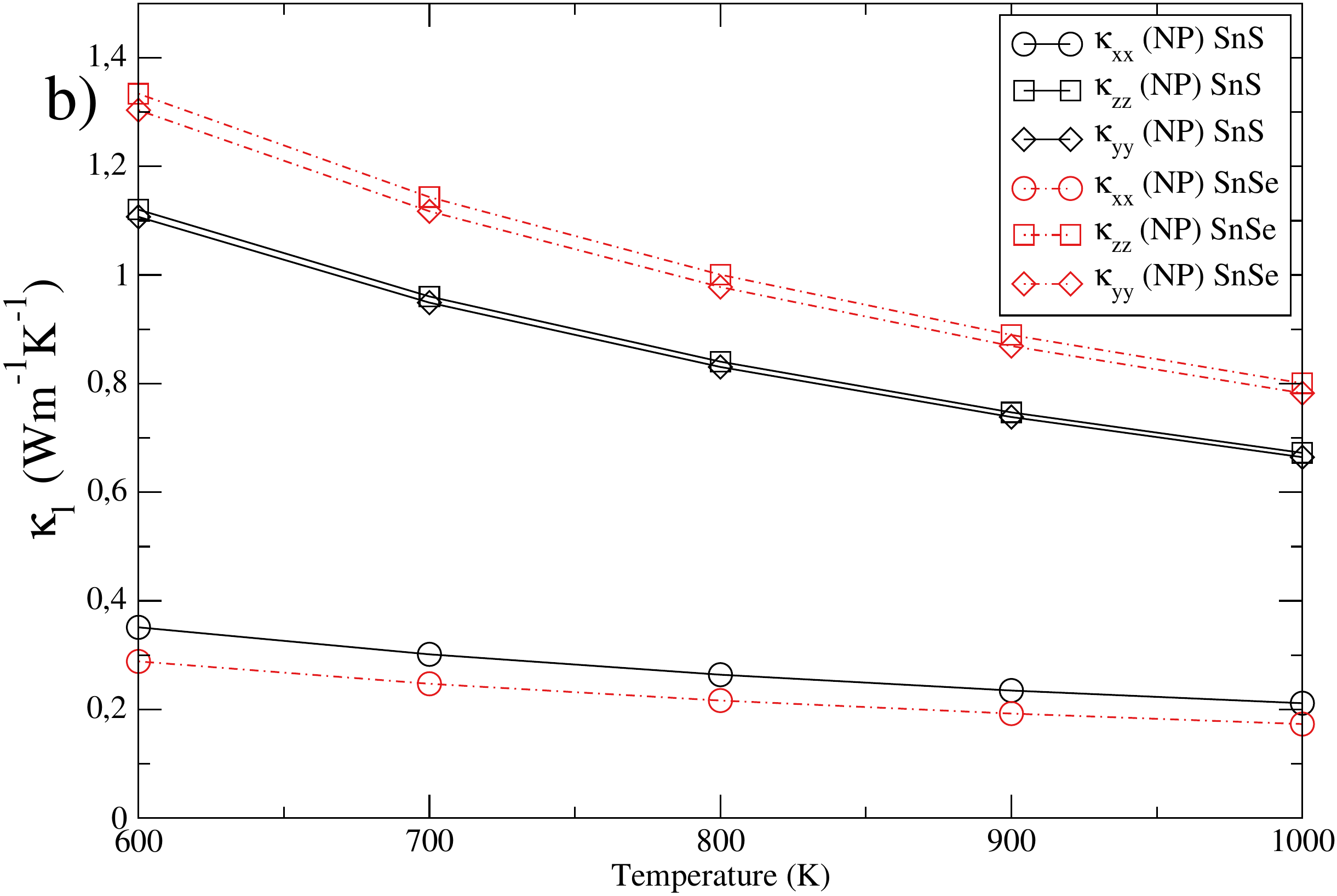}
\caption{a) Lattice thermal conductivity of $Cmcm$ SnS calculated within non-perturbative (NP) and perturbative (P) approaches. We have used $\tilde{\Omega}_{\mu}(\mathbf{q})$ at $800$ K for both and $\overset{(3)}{\boldsymbol{\Phi}}$ 
at $800$ K for the non-perturbative calculation in both cases. Calculations are within the LDA. b) Lattice thermal conductivity of $Cmcm$ SnS and SnSe calculated within the non-perturbative (NP) approach.} 
\label{tk}
\end{figure}

In Figure \ref{tk} (a) we show the lattice thermal conductivity of $Cmcm$ SnS as a function of temperature calculated using $\overset{(3)}{\boldsymbol{\Phi}}$ and $\overset{(3)}{\boldsymbol{\phi}}$ for solving the BTE within 
the SMA. We recall that $\overset{(3)}{\boldsymbol{\Phi}}$ are non-perturbative third-order force-constants calculated using Eq. \ref{fcnp} and $\overset{(3)}{\boldsymbol{\phi}}$ are perturbative third-order force-constants 
calculated using Eq. \ref{fcp}. In Figure \ref{tk} (b) we show the lattice thermal conductivities of $Cmcm$ SnS and SnSe using $\overset{(3)}{\boldsymbol{\Phi}}$.
We can see that the non-perturbative calculation using $\overset{(3)}{\boldsymbol{\Phi}}$ is lower than the perturbative one using $\overset{(3)}{\boldsymbol{\phi}}$ for the three Cartesian directions. This result makes clear that 
the non-perturbative anharmonicity is very important to calculate the thermal conductivity in this kind of thermoelectric materials. By looking at the values of the lattice thermal conductivity we can see that both materials show 
very similar ultralow values, below $\approx 1.0$ Wm$^{-1}$K$^{-1}$. In-plane results are slightly higher for SnSe and out-of-plane calculations higher for SnS, in agreements with another calculation\cite{guo2015first} where the 
thermal conductivities of SnS and SnSe for the low-temperature $Pnma$ phase are calculated with harmonic phonons and perturbative third-order force-constants. Theoretical calculations following the same procedure also show that the 
thermal conductivities of $Pnma$ SnSe and SnS are very similar\cite{skelton2017lattice,skelton2016anharmonicity}, in agreement with experiments\cite{zhao2014ultralow,tan2014thermoelectrics}. Our calculations confirm that in the 
high-temperature $Cmcm$ phase the thermal conductivity of these two compounds is also very similar. Both materials show a clear anisotropy between in-plane and out-of-plane calculations in agreement with experimental 
results\cite{ibrahim2017reinvestigation} for the low-temperature phase close to the phase transition. 

\section{Conclusions}
\label{conclusions}

In conclusion, we have calculated the electronic and vibrational transport properties of $Cmcm$ SnS using first-principles calculations. We have seen that the electronic transport properties of SnS and SnSe are comparable and that a 
similar power factor is expected for these isoelectronic and isostructural compounds. As in the case of SnSe, SnS suffers a second-order phase transition from the $Cmcm$ to the $Pnma$ phase driven by the collapse of a zone border 
phonon. We have also seen that SnS shows a strongly anharmonic phonon spectrum. Many phonon modes have a very large linewidth and show non-Lorentzian profiles in the spectral function. Finally, we have calculated the lattice thermal 
conductivity of $Cmcm$ SnS and we have seen that nonperturbative anharmonicity substantially corrects the third order force-constants. The thermal conductivity of both materials is very similar and 
ultralow. Therefore, by comparing the electronic and vibrational transport properties of SnS and SnSe in the $Cmcm$ high-temperature phase, we conclude both should be good thermoelectrics. Thus, we suggest that SnS may be an 
earth-abundant very efficient high-temperature thermoelectric material. This work motivates more experimental effort in this regime for its characterization. \\

\section{ACKNOWLEDGMENTS}

Financial support was provided by the Spanish Ministry of Economy and Competitiveness (FIS2016-76617-P); and the Department of Education, Universities and Research of the Basque Government and the University of the Basque 
Country (IT756-13). U.A. is also thankful to the Material Physics Center for a predoctoral fellowship. Computer facilities were provided by the Donostia International Physics Center (DIPC), PRACE (2017174186) and 
DARI (A0050901202).

\bibliography{bibliografia}% Produces the bibliography via BibTeX.

\end{document}